\label{} 

\documentclass[preprint,aps]{revtex4}
\usepackage{amsmath,amssymb,amsfonts,times,graphicx,color}

\usepackage{graphicx}
\usepackage{dcolumn}
\usepackage{bm}

\def\barray{\begin{eqnarray}}
\def\earray{\end{eqnarray}}
\def\beq{\begin{equation}}
\def\eeq{\end{equation}}

\begin{document}

\preprint{}

\title{Information Geometry of Entanglement Renormalization for free Quantum Fields} 

\author{J. Molina-Vilaplana}
\affiliation{Universidad Polit\'ecnica de Cartagena. C/Dr Fleming S/N. 30202. Cartagena. Spain}

\email{javi.molina@upct.es}


\begin{abstract}
We provide an explicit connection between the differential generation of entanglement entropy in a tensor network representation of the ground states of two  field theories, and a geometric description of these states based on the Fisher information metric. We show how the geometrical description remains invariant despite there is an irreducible gauge freedom in the definition of the tensor network. The results might help to understand how spacetimes may emerge from distributions of quantum states, or more concretely, from the structure of the quantum entanglement concomitant to those distributions.
\end{abstract}

                                                            
\maketitle

\section{Introduction}
\label{intro}
Recently, it has been proposed that the structure of spacetime in gravitational theories may inextricably be related with the entanglement structure of some fundamental degrees of freedom \citep{ent_geom1, ent_geom2}. The holographic formula of the entanglement entropy \cite{ryutak}
\beq
\label{eq0.0}
S(A)=\frac{1}{4\, G_N^{(d+2)}}\, {\rm Area}(\gamma_A),
\eeq
which provides a prescription to quantify the entanglement entropy $S(A)$ of a region $A$ in a $(d+1)$-QFT which admits a $(d+2)$-gravity dual given by the AdS/CFT correspondence  \cite{adscftbib2,adscftbib3,adscftbib4}, happens to be a first manifestation of this conjecture. Here, $\gamma_A$ is the codimension-2 static minimal surface in AdS$_{(d+2)}$ whose boundary and area are given by $\partial A$ and ${\rm Area}(\gamma_A)$ respectively. In this context, it has been shown that entanglement entropy obeys a \textit{first law}, an exact quantum generalization of the ordinary first law of thermodynamics. Thus, in any CFT with a semiclassical holographic dual, this first law has an interpretation in the dual gravitational theory as a constraint on the spacetimes dual to CFT states. Based on the Ryu-Takayanagi proposal and this first law of entanglement, it is possible to extract the dynamics of the emergent space-time at linearized level (for certain entangling regions). This approach has been pioneered by authors in \cite{firstlaw1} and followed much more precisely in \cite{firstlaw2}. 

Another proposal to understand the relationship between the structure of quantum entanglement and an emergent space-time has recently emerged. Following suggestions in \cite{ent_geom1}, authors in \cite{ent_geom2} proposed that any two entangled quantum systems may admit a dual gravitational description given by a non-transversable wormhole geometry. In most cases, the wormhole duals are hard-to-describe strongly fluctuating quantum mechanical geometries, but in some cases, the wormhole duals may possess a smooth Riemannian geometry. 

In addition, using MERA (multi-scale entanglement renormalization ansatz, \cite{vidalmera}) tensor network representations (particularly its continuous version, cMERA, \cite{cMERA}), a gravitational-like geometric description of some relevant states in quantum many body systems and field theories have been provided \cite{swingle, taka1}. We refer to \cite{prog1,prog2,prog3,prog4,prog6} for more recent progress on this subject. However, the relation between the entanglement structure underlying the tensor network construction and its geometrical description has been established only on qualitative grounds. 

In this work, we find that the square of the differential generation of entanglement entropy along the renormalization group flow implemented by cMERA, amounts to a geometrical description of the ground states of two (1+1)-dimensional QFTs given in terms of the Fisher information metric. Instead of focusing on a geometric description of the entangled subsystems, we will carry out our analysis by considering the entanglement between the left and right-moving modes of the QFT. Furthermore, it will be shown how the emergent geometrical description of the state, remains invariant despite there is an irreducible gauge freedom in the definition of the cMERA network.

The paper is structured as follows: in Section 2, we briefly review the formalism of entanglement renormalization for continuous quantum systems (cMERA), especially focusing on the coherent state formulation for free Gaussian theories. In any case, we refer to \cite{cMERA, taka1} for more extensive treatments and presentations of the topic. Section 3 is devoted to the computation of the left-right entanglement entropy and its flow along the cMERA renormalization group process. At the end of the section, we provide some hints on the emergent geometrical interpretation of the cMERA differential entanglement flow in terms of the relative entropy, a measure of distinguishability between quantum states. In Section 4, we comment on the emergent geometry describing the cMERA renormalization and its relation with the flow of left-right entanglement computed in Section 3. It is shown how this geometric description remains invariant under some class of local gauge transformations defined along the cMERA renormalization flow. Finally we summarize our results and suggest some issues to be investigated in the future.
 
\section{Entanglement Renormalization for QFT}
\label{cMERA}
Entanglement renormalization (MERA) is a real-space renormalization group formulation on the quantum state (instead of the Wilsonian RG scheme) \cite{vidalmera, cMERA}. MERA represents the wavefunction of the system at each relevant length scale $u$ of the system. By convention, $u=0$ refers to the state description at short lenghts (UV-state $|\Psi_{UV}\rangle$). Starting from it, (in principle, this amounts to a highly entangled state), each scale $u$ of MERA performs a renormalization transformation in which, prior to coarse graining the effective degrees of freedom at that scale, the short range entanglement between them is removed through the  action of an unitary transformation called {\em disentangler}. Thus iteratively, MERA removes the quantum correlations between small adjacent regions of space at each length scale. This RG procedure is applied arbitrarily many times until one reaches the IR-state $|\Psi_{IR}\rangle$\footnote{The state $|\Psi_{IR}\rangle$ has no real space entanglement, i.e, is a completely unentangled state in case of massive theories. When considering a massles CFT, this state coincides with the vacuum $|0\rangle$ of the theory which is an entangled state.}. Namely, the procedure may be run backwards so, starting from $|\Psi_{IR}\rangle$, one unitarily adds entanglement at each length scale until the correct $|\Psi_{UV}\rangle$ is generated.

To be precise, let us consider the state $|\Psi(u)\rangle$ obtained by adding entanglement between modes of momentum $k \leq \Lambda e^{-u}$ to the unentangled state $|\Psi_{IR}\rangle$, 
\beq
\label{eq1.1}
|\Psi(u)\rangle = P\, e^{-i\int_{u_{IR}}^{u} d\hat{u}(K(\hat{u}) + L)}\, |\Psi_{IR}\rangle,
\eeq
where $P$ is a path ordering symbol which allocates operators with bigger $u$ to the right and $\Lambda$ is the UV momentum cut-off. The operator $K(\hat{u})$ generates the entanglement along the cMERA flow from $u_{IR}$ to a given $u$. It reads as,
\beq
\label{eq1.2}
K(\hat{u})=\int d^d k\, \Gamma(k/\Lambda)\, g(\hat{u},k)\, \mathcal{O}_k,
\eeq
where $\mathcal{O}_k$ is an operator acting at the energy scale given by $k$ and $\Gamma(x)=1$ for $0<x<1$ and zero otherwise. The function $g(\hat{u},k)$ is model/state dependent and gives the strenght of the entangling process at a given scale. The operator $L$ corresponds to the coarse-graining process \cite{cMERA,taka1}. In this paper we mainly focus on the entangling process, thereby, to get rid of the $L$ process in our analysis, we proceed by rescaling the cMERA states as,
\beq
\label{eq1.3}
|\widetilde{\Psi}(u)\rangle = P\, e^{-i\int_{u_{IR}}^{u} d\hat{u}\, \widetilde{K}(\hat{u})}\, |\Psi_{IR}\rangle.
\eeq
Here, the \textit{entangler} operator is given in the \textit{interaction} picture $ \widetilde{K}(\hat{u}) = e^{-i\hat{u}L}\, K(\hat{u})\, e^{i\hat{u}L}$, and reads as,
\beq
\label{eq1.4}
 \widetilde{K}(\hat{u}) = \int d^d k\, \Gamma(k\, e^{\hat{u}} /\Lambda)\, g(\hat{u},k\, e^{\hat{u}})\, \widetilde{\mathcal{O}}_k,
\eeq
with $ \widetilde{\mathcal{O}}_k = e^{-i\hat{u}L}\, \mathcal{O}_k\, e^{i\hat{u}L}$. 

In this paper, we will consider two examples of free fields in (1+1) dimensions, namely the free massive boson and a free massive Dirac fermion. For the free boson theory with action,
\beq
\label{eq1.5}
S_B=\int dt dx \left[\left(\partial_t\, \phi \right)^2 + \left(\partial_x\, \phi \right)^2 -m^2 \phi^2 \right], 
\eeq 
one has,
\beq
\label{eq1.6}
 \widetilde{K}_B(\hat{u}) = i\int dk\, \left( g_k^B(\hat{u})\, a^{\dagger}_k\, a^{\dagger}_{-k} - g_k^{B}(\hat{u})^{*}\, a_k\, a_{-k}\right),
\eeq
where $g_k^B(\hat{u})= \Gamma(k e^{\hat{u}}/\Lambda)\, g^B(\hat{u},k)$ and $a^{\dagger}_k, a_k$ are the creation and anihilation operators of the field mode with momentum $k$ such that, if $|0\rangle_B$ is the vacuum state of the theory then, $a_k |0\rangle_B = a_{-k} |0\rangle_B=0$. 

The free Dirac fermion theory is given by the action,
\beq
\label{eq1.7}
S_F=\int dt dx \left[i\overline{\psi}\left(\gamma^t\partial_t\, + \gamma^x\partial_x \right)\psi  -m \overline{\psi}\psi \right], 
\eeq 
where $\psi$ is a two component complex fermion with $\gamma^t=\sigma_3,\, \gamma^x=i\sigma_2$ and $\overline{\psi}=\psi^{\dagger}\gamma^t$. The \textit{entangler} operator in this case reads as,
\beq
\label{eq1.8}
 \widetilde{K}_F(\hat{u}) = i\int dk\, \left( g_k^F(\hat{u})\, c^{\dagger}_k\, d_{k} +g_k^{F}(\hat{u})^{*}\, c_k\, d^{\dagger}_{k}\right),
\eeq
where  $g_k^F(\hat{u})=(k\, e^{u}/\Lambda)\, \Gamma(k e^{\hat{u}}/\Lambda)\, g^F(\hat{u},k)$ and $c_k,\, d^{\dagger}_{k}$ are the anihilation operators for field modes of each component (particles and anti-particles) such that \cite{taka1},
\beq
\label{eq1.9}
c_k |0\rangle_F = d^{\dagger}_k |0\rangle_F=0.
\eeq

In the following we shall omit the subscripts $(B,F)$ while it will be clear to which case we are referring.
\subsection*{Coherent State description of cMERA}
In the bosonic theory, the state in eq.(\ref{eq1.3}) may equivalently be written as \cite{taka1},
\beq
\label{eq1.10}
|\widetilde{\Psi}(u)\rangle =  \mathcal{N}\, \exp\left[ \int dk\, \Phi_k(u)\,a^{\dagger}_k\, a^{\dagger}_{-k}\right] |0\rangle = \mathcal{N}\, \prod_{k} \exp\left[\Phi_k(u)\, a^{\dagger}_k\, a^{\dagger}_{-k}\right] |0\rangle,
\eeq
where,
\beq
\label{eq1.10i}
\Phi_k(u) = \int_{0}^{u}  g_k(\hat{u})\, d\hat{u}.
\eeq   
The state is normalized by taking $\mathcal{N}=\exp\left[-1/2\, \int dk\, |\Phi_k(u)|^2 \right]$. This state is a Gaussian coherent state annihilated by the operator,
\beq
\label{eq1.11}
b_k(u)=A_k(u)\, a_k + B_k(u)\,  a^{\dagger}_{-k},
\eeq
i.e, $b_k(u)|\widetilde{\Psi}(u)\rangle = 0$ with $|A_k(u)|^2-|B_k(u)|^2=1$. Eq.(\ref{eq1.11}) amounts to a scale-dependent Bogoliubov transformation whose model dependent coefficients are given by \cite{cMERA},
\barray
\label{eq1.12}
A_k(u)& = &\cosh \Phi_k(u)\, \alpha_k -\sinh \Phi_k(u)\, \beta_k \\\nonumber
B_k(u)& = &-\sinh \Phi_k(u)\, \alpha_k +\cosh \Phi_k(u)\, \beta_k,
\earray
with $\alpha_k \equiv A_k(u_{IR}),\, \beta_k \equiv B_k(u_{IR})$. Thus, the state $|\Psi_{IR}\rangle$ is defined as,
\beq
\label{eq1.13}
(\alpha_k\, a_k +\beta_k\, a^{\dagger}_{-k})|\Psi_{IR}\rangle=0.
\eeq

In the fermionic theory, $|\widetilde{\Psi}(u)\rangle$ reads as,
\beq
\label{eq1.14}
|\widetilde{\Psi}(u)\rangle =  \mathcal{N}\, \exp\left[ \int dk\, \Phi_k(u)\,c^{\dagger}_k\, d_{k}\right] |0\rangle =  \mathcal{N}\, \prod_{k} \exp\left[\Phi_k(u)\, c^{\dagger}_k\, d_{k}\right] |0\rangle,
\eeq
where again, $\Phi_k(u) = \int_{0}^{u}  g_k(\hat{u})\, d\hat{u}$ and the state is normalized by $\mathcal{N}=\exp\left[-1/2\, \int dk\, |\Phi_k(u)|^2 \right]$. Eq.(\ref{eq1.14}) is a \textit{displaced} vacuum coherent state which is annihilated by the operator,
\beq
\label{eq1.15}
\psi_k(u)=A_k(u)\, c_k + B_k(u)\,  d^{\dagger}_{k},
\eeq
i.e, $\psi_k(u)|\widetilde{\Psi}(u)\rangle = 0$ with coefficients,
\barray
\label{eq1.16}
A_k(u)& = &\cos \Phi_k(u)\, \alpha_k + \sin \Phi_k(u)\, \beta_k \\\nonumber
B_k(u)& = &-\sin \Phi_k(u)\, \alpha_k +\cos \Phi_k(u)\, \beta_k,
\earray
such that $|A_k(u)|^2+|B_k(u)|^2=1$, and
\beq
\label{eq1.17}
(\alpha_k\, c_k +\beta_k\, d^{\dagger}_{k})|\Psi_{IR}\rangle=0.
\eeq

In this framework, the entangling operation of cMERA in the free theories under consideration amounts to a sequential generation of a set of coherent states $|\widetilde{\Psi}(u)\rangle$ defined through eqs.(\ref{eq1.10}), (\ref{eq1.14}). In both cases, $|\widetilde{\Psi}(u)\rangle$ is an non-entangled vacuum for the Bogoliubov-quasiparticles at that scale, while as  \textit{displaced} vacuum states, they are highly entangled relative to any state defined on a higher scale of cMERA. 

\section{Entanglement flow in MERA}
\label{Entanglement}
In this section, we quantify the entanglement flow required to generate $|\widetilde{\Psi}(u)\rangle$ starting from $|\Psi_{IR}\rangle$. Let us first consider the bosonic case by writing the state in eq.(\ref{eq1.10}) as a superposition of Fock states,
\beq
\label{eq2.1}
|\widetilde{\Psi}(u)\rangle= \prod_{k}\, \sum_{n=0}^{\infty}\, c_n^k\, |n_k,\, n_{-k}\rangle =  \prod_{k} |\Psi_k(u)\rangle ,
\eeq
where
\beq
\label{eq2.1i}
|\Psi_k(u)\rangle = \sum_{n=0}^{\infty}\, c_n^k\, |n_k,\, n_{-k}\rangle,
\eeq
Fock states $|n_k,\, n_{-k}\rangle \propto (a_k^{\dagger})^n\, (a_{-k}^{\dagger})^n\, |0\rangle$ and,
\beq
\label{eq2.2}
c_n^k = \gamma_k(u)^{n/2}\, \sqrt{1-\gamma_k(u)}, \quad \gamma_k(u) = \left[ \frac{B_k(u)}{A_k(u)}\right] ^2.
\eeq
Here, $A_k(u)$ and $B_k(u)$ are those in eq.(\ref{eq1.12}). The total amount of entanglement generated between all the modes with opposite momenta ($|k| \leq \Lambda e^{-u}$) when creating $|\widetilde{\Psi}(u)\rangle$ from $|\Psi_{IR}\rangle$ amounts to the von Neumann entropy of
\barray
\label{eq2.3}
\rho(u)={\rm Tr}_{[-k]}\left(|\widetilde{\Psi}(u)\rangle \langle \widetilde{\Psi}(u)|\right)  = \prod_{k} \sum_{n=0}^{\infty}\, \vert c^k_n\vert^2\,  |n_k\rangle \langle n_k| =\prod_{k} \sum_{n=0}^{\infty}\, \gamma^n\, (1-\gamma)\,  |n_k\rangle \langle n_k|,  
\earray
where $\gamma \equiv \gamma_k(u)$. In a free theory where all modes are decoupled, the entanglement entropy $\mathcal{S}(u)$ can be written as,
\beq
\label{eq2.4}
\mathcal{S}(u) = -\int_{0}^{\Lambda e^{-u}}  dk\, {\rm Tr}\left[\, \rho_k(u)\log \rho_k(u)\, \right],
\eeq
with
\beq
\label{eq2.4i}
\rho_k(u)= \sum_{n=0}^{\infty}\, \gamma_k(u)^n\, (1-\gamma_k(u))\,  |n_k\rangle \langle n_k|.
\eeq
Thus, it is possible to carry out the analysis only focusing on the entanglement generated between two modes with opposite momenta (left-right moving modes), i.e,
\beq
\label{eq2.4ii}
S_k(u)=-{\rm Tr}\left[\, \rho_k(u)\log \rho_k(u)\, \right]. 
\eeq
A standard calculation for this entropy yields \cite{bombelli},
\beq
\label{eq2.5}
S_k(u)=\frac{\gamma_k(u)}{\gamma_k(u)-1}\, \log \gamma_k(u) - \log(1-\gamma_k(u)).
\eeq

On the other hand, the entanglement flow in the process amounts to quantify how much entanglement is added at each infinitesimal cMERA layer. By differentiating eq.(\ref{eq2.5}) wrt $u$ and noticing that $\partial_u \Phi_k(u) = g_k(u)$, one obtains,
\beq
\label{eq2.6}
\partial_u S_k(u)=\left[ \frac{2\sqrt{\gamma_k(u)}}{(1-\gamma_k(u))}\, \log \gamma_k(u)\right] \, g_k(u),
\eeq
which explicitly relates the rate of entanglement generation with the stregth of the \textit{entangling} operation $g_k(u)$. When $\gamma_k(u) \sim 1$, the factor $\left( 2\sqrt{\gamma_k(u)}/(1-\gamma_k(u))\right) \, \log \gamma_k(u) \approx -(1+\gamma_k(u)) \approx -2$. This allows to write,
\beq
\label{eq2.7}
g_k(u)\approx -\frac{1}{2}\, \partial_u S_k(u) .
\eeq
Figure 1 illustrates this relation for the ground state of a free scalar theory with mass $m$. In this case, by variationally minimizing the energy density $E=\langle \Psi_{IR}|\mathit{H}(u_{IR})|\Psi_{IR}\rangle$ for $k<\Lambda\, e^{-u}$, one obtains \cite{cMERA,taka1},
\beq
\label{eq2.8}
g_k(u)=g(u)=-\frac{1}{2}\, \frac{e^{-2 u}}{e^{-2u}+m^2/\Lambda^2}.
\eeq
where $\mathit{H}$ is the Hamiltonian of the system. 

\begin{figure}[t]
\label{entopy_cMERA}
\includegraphics[width=4.0 in]{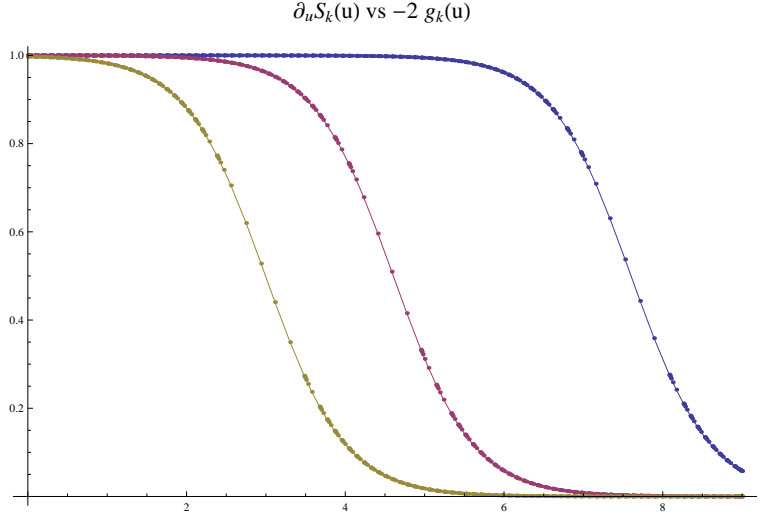}
 \centering \caption{Entropy rate $\partial_u S_k(u)$ (dots) vs $-2g_k(u)$ (continuous line) for three different masses $m= 0.05, 1, 5$ of the free boson. The plot has been created taking $k=0.005$ and $\Lambda = 100$.  $\partial_u S_k(u)$ is computed by numerically differentiating values of $S_k(u)$ obtained through eq.(\ref{eq2.5}). The cMERA scale $u$ runs from the 0 to 9. For each case, the rate $\partial_u S_k(u)$ vanishes at a different scale $u_{IR}$ which increases as $m$ decreases. This scale indicates that the renormalization process has reached the state $|\Psi_{IR}\rangle$.}
\end{figure}

In the fermionic case, as $\psi_k(u)|\widetilde{\Psi}(u)\rangle=0$, and taking into account eq.(\ref{eq1.15}), the state in eq.(\ref{eq1.14}) can be written as,
\barray
\label{eq2.9}
|\widetilde{\Psi}(u)\rangle & = & \mathcal{N}\, \prod_k\, \left( 1 + \gamma_k^{1/2}(u)\, c_k^{\dagger}\, d_k\right) |0\rangle \nonumber \\
&=&  \mathcal{N}\, \prod_k\, \left( |0_c, 0_d\rangle_k + \gamma_k^{1/2}(u)|1_c, 1_d\rangle_k\right)= \mathcal{N}\, \prod_k |\Psi_k(u)\rangle. 
\earray
Here, $\gamma_k(u)=\left[B_k(u)/A_k(u)\right]^2 $ with $A_k(u)$ and $B_k(u)$  given by eq.(\ref{eq1.16}). The states $|0_c, 0_d\rangle_k$ and $|1_c, 1_d\rangle_k$ refer to fermionic Fock states with no $c$-particles and $d$-antiparticles of momentum $k$ and one $c$-particle and one $d$-antiparticle of momentum $k$ respectively; $|0\rangle \equiv \prod_k\, |0_c, 0_d\rangle_k$ and
\beq
\label{eq2.10}
|\Psi_k(u)\rangle = \frac{1}{\sqrt{1+\gamma_k(u)}}\left( |0_c, 0_d\rangle_k + \gamma_k(u)^{1/2}|1_c, 1_d\rangle_k\right).
\eeq
In like manner as before, we proceed by focusing on the left-right entanglement of the state $|\Psi_k(u)\rangle$. This amounts to the entanglement between a $c$-mode and a $d$-mode given by the entropy $S_k(u)=-{\rm Tr}\left[\, \rho_k(u)\log \rho_k(u)\, \right]$, where the reduced density matrix $\rho_k(u)$ reads as,
\barray
\label{eq2.11}
 \rho_k(u) = {\rm Tr}_{[d\, ]}\left(|\Psi_k(u)\rangle \langle \Psi_k(u)| \right)  = \left( \begin{array}{cc}
1/(1 + \gamma_k(u)) & 0 \\  
0 & \gamma_k(u)/(1 + \gamma_k(u)) \end{array} \right).
\earray
Then, a straightforward calculation yields,
\beq
\label{eq2.12}
S_k(u) = \log (1 + \gamma_k(u))-\frac{\gamma_k(u)}{(1+\gamma_k(u))}\, \log \gamma_k(u).
\eeq

To obtain the rate of entanglement generation along the cMERA flow in the free fermion theory, one simply differentiates eq.(\ref{eq2.12}). The entanglement flow, as in the bosonic case, results proportional to the strength of the entangling operation and can be written as,
\beq
\label{eq2.13}
\partial_u\, S_k(u) =  \left[\frac{2\sqrt{\gamma_k(u)}}{(1+\gamma_k(u))}\log \gamma_k(u) \right]\, g_k(u).
\eeq

Both eq.(\ref{eq2.7}) and eq.(\ref{eq2.13}) are major results of this work and, as it will be shown below, they shall allow to write explicit formulas linking the rate of entanglement generation in cMERA flows with the geometric descriptions of the process proposed in \cite{taka1}.

\subsection*{Relative Entropy, Fisher Metric and cMERA Coherent States}
\label{relative_ent}
A measure of distinguishability between the quantum probability distributions defined by $\bar{\rho}\equiv\rho_k\left( u+du,\left\lbrace \bar{\gamma}\right\rbrace \right) $ and $\rho\equiv\rho_k\left( u,\left\lbrace \gamma\right\rbrace \right) $ has been computed in terms of the relative entropy between them \cite{vedral}
\beq
\label{eq2.14}
S(\, \bar{\rho}\, ||\, \rho\, )={\rm Tr}\left[\, \bar{\rho}\log\bar{\rho}-\bar{\rho}\log \rho \right], 
\eeq
where it has been assumed that $\bar{\gamma}\equiv \gamma(u+du)\approx \gamma(u)+ \partial_u\, \gamma(u)\, du$. In the bosonic theory, the computation yields,
\barray
\label{eq2.15}
S(\, \bar{\rho}\, ||\, \rho\, )  =  \frac{\bar{\gamma}}{(1-\bar{\gamma})}\log\frac{\bar{\gamma}}{\gamma} + \log\frac{(1-\bar{\gamma})}{(1-\gamma)}= 4\, g_k(u)^2\, du^2.
\earray
In like manner, for the fermionic case one obtains,
\barray
\label{eq2.16}
S(\, \bar{\rho}\, ||\, \rho\, )=  \frac{\bar{\gamma}}{(1+\bar{\gamma})}\log\frac{\bar{\gamma}}{\gamma} - \log\frac{(1+\bar{\gamma})}{(1+\gamma)}= 4\, g_k(u)^2\, du^2.
\earray

In information geometry, the Fisher information metric \cite{amari} is a Riemannian metric defined on a smooth statistical manifold, i.e., a smooth manifold whose points are probability distributions defined on a common probability space. The metric measures the informational difference between those points (distributions) and  amounts to the infinitesimal form of the relative entropy. The computations above, addressed the case in which these probability distributions correspond to the reduced density matrices $\bar{\rho}$ and $\rho$ of two infinitesimally displaced cMERA quantum states. 

At this point, let us comment on the intimate relationship between the cMERA states and the symmetry group underlying the entanglement renormalization group flow. First, we consider a set of operators $\lbrace \mathcal{T}_i\rbrace$ with conmutators,
\beq
\label{eq2.17}
\left[ \mathcal{T}_i,  \mathcal{T}_j\right]=c_{ij}^k\, \mathcal{T}_k.
\eeq
The set $\lbrace \mathcal{T}_i\rbrace$ span an algebra $\mathfrak{g}$, with $c_{ij}^k$ the structure constants of $\mathfrak{g}$. If $\mathfrak{g}$ constitutes a semisimple Lie algebra, it is rather convenient to express $\lbrace \mathcal{T}_i\rbrace$ in terms of the Cartan basis $\lbrace N_i, \, E_{\alpha},\, E_{-\alpha}\equiv E_{\alpha}^{\dagger}\rbrace$:
\barray
\label{eq2.18}
\left[ H_i,  H_j\right]&=&0, \quad \left[ H_i,  E_{\alpha}\right]=\alpha_i\, E_{\alpha}, \\ \nonumber
\left[ E_{\alpha},  E_{-\alpha}\right]&=&\alpha^{i}\, H_{i} \quad \left[ E_{\alpha},  E_{\beta}\right]=N_{\alpha\, \beta}\,  E_{\alpha + \beta}.
\earray
For such a closed set of operators, the states of the associated quantum theory belong to a Hilbert space $\mathcal{H}$ which amounts to a representation of $\mathfrak{g}$. Namely, if $\mathbf{G}$ is the covering group of $\mathfrak{g}$, the Hilbert space $\mathcal{H}$ amounts to an irreducible unitary representation of $\mathbf{G}$. Thus, it is possible to take a normalized state $|\psi_0\rangle \in \mathcal{H}$ as a fixed state, such that, a coherent state can be generated by an element $g \in \mathbf{G}$ as,
\beq
\label{eq2.19}
|\psi_g\rangle_{\mathbf{G}}=g\, |\psi_0\rangle.
\eeq
The element $g \in \mathbf{G}$ may uniquely decomposed into $g=k \cdot h$, with $h \in \mathbf{H}$, and $\mathbf{H}$ the maximum subgroup of $\mathbf{G}$ whose action, leaves invariant the referent state up to a phase,
\beq
\label{eq2.20}
|\psi_h\rangle=h\, |\psi_0\rangle = e^{i\, \eta} |\psi_0\rangle.
\eeq
On the other hand, $k$ is on the coset space $\mathbf{G}/\mathbf{H}$ and provided that $\mathbf{G}$ is a semisimple Lie group, it can be written as an operator wich gives a coset representation of $\mathbf{G}/\mathbf{H}$, called \textit{displacement operator} $D(\Phi)$. In this sense,
\beq
\label{eq2.21}
|\psi_g\rangle_{\mathbf{G}}=D(\Phi)\, e^{i\, \eta} |\psi_0\rangle \equiv e^{i\, \eta}\, |\Psi(\Phi)\rangle.
\eeq
The state $|\Psi(\Phi)\rangle$ is known as the coherent state of $\mathbf{G}/\mathbf{H}$ and can be written as \cite{perelmonov}:
\beq
\label{eq2.22}
|\Psi(\Phi)\rangle=\mathcal{N}(\Phi)\, \exp \left( \sum_{\alpha} \Phi_{\alpha}\, E_{\alpha} \right) |\psi_0\rangle,
\eeq
with $\mathcal{N}(\Phi)$ a normalization constant. These states satisfy,
\beq
\label{eq2.23}
\int\, d\mu(\Phi)\, |\Psi(\Phi)\rangle \langle \Psi(\Phi)| = I,
\eeq
where $d\mu(\Phi)$ is the $\mathbf{G}$-invariant Haar measure on $\mathbf{G}/\mathbf{H}$. In addition, and of great interest to us, each one of these states are one-to-one corresponding to the points in the coset $\mathbf{G}/\mathbf{H}$ manifold except for some singular points. As a result, the states $|\Psi(\Phi)\rangle$ are embeded into a topologically nontrivial space.

Regarding how the cMERA renormalization group flow is expressed in terms of the coherent states $|\widetilde{\Psi}(u)\rangle \equiv |\widetilde{\Psi}(\Phi)\rangle$, we note that these states are obtained through the displacement operator (in the bosonic theory) $D(\Phi)\in SU(1,1)/U(1)$ \cite{perelmonov},
\beq
\label{eq2.24}
D(\Phi) = \exp\left[\int dk\,\left( \Phi_k(u)\, a^{\dagger}_k\, a^{\dagger}_{-k} - \Phi_k(u)^{*}\, a_k\, a_{-k}\right)\right],
\eeq
acting on the vacuum state $|0\rangle$ (which amounts to the reference state $|\psi_0\rangle$). Remarkably, the group manifold $SU(1,1)/U(1)$ corresponds to a 2-dimensional hyperbolic space. In other words, each cMERA state $|\widetilde{\Psi}(u)\rangle$ (a quantum probability distribution) corresponds to a point on a two dimensional hyperbolic space.  Thus, it is reasonable to argue that once provided a suitable measure of the distance between the states $|\widetilde{\Psi}(u)\rangle$, then a geometric description of the cMERA renormalization flow should correspond to the metric of a two dimensional AdS space. The results on the relative entropy given above also indicate that it would be possible to relate this metric with the strength of the disentangling operation $g_k(u)$ and therefore, through eq.(\ref{eq2.7}) and eq.(\ref{eq2.13}), with the differential generation of entanglement entropy along the renormalization group flow.

\section{Emergent Geometry and Entanglement}
\label{geometry:cMERA}
In \cite{swingle}, it has been conjectured that, from the entanglement structure of an static (1+1) wavefunction represented by an \textit{entanglement renormalization} tensor network, one may define a higher dimensional geometry in which, apart from the coordinate $x$, it is reasonable to define a "radial" coordinate $u$ which accounts for the hierarchy of scales. Namely, in the AdS/CFT, it is widely accepted that the holographic radial dimension corresponds to the length scale of the renormalization group flow, whereupon it is natural to identify the length scale $u$ with the radial direction of a dual geometric description of the quantum state \cite{swingle}. This conjecture has been qualitatively confirmed by comparing how entanglement entropy is computed in MERA tensor networks and in the AdS/CFT correspondence \cite{ryutak}. The geometry emerging at the critical point is the hyperbolic AdS spacetime. For more generic static cMERA states, it is hypothesized that the metric must be an asymptotically AdS geometry given by,
\beq
\label{eq3.0}
ds^2 = g_{uu}\, du^2 + \Lambda^2\, e^{-2 u}dx^2.
\eeq

Recalling our latest comments on the previous section, one may put into context this interpretation by noting that a geometric description of cMERA similar to the latter can be defined by only invoking to information theoretic concepts, without any need to assume the existence of a AdS dual. To this end, we note  that in \cite{taka1}, authors obtained the Fisher information metric which measures distances between the cMERA states $\lbrace |\widetilde{\Psi}(u)\rangle\,/ \, u \in \left[\, 0,\, u_{IR}\, \right]  \rbrace $. The distance measure $\mathcal{D}\left[\widetilde{\Psi}(u),\, \widetilde{\Psi}(u+du) \right] $ in the Hilbert space spanned by these cMERA coherent states, was chosen to be the Hilbert-Schmidt distance,
\beq
\label{eq3.1}
\mathcal{D}^{2}_{HS}\left[\widetilde{\Psi}(u),\, \widetilde{\Psi}(u+du) \right]= 1-|\langle \widetilde{\Psi}(u)|\widetilde{\Psi}(u+du)\rangle|^2 .
\eeq
With this election, the proposal for the $g_{uu}$ component of the metric reads as,
\beq
\label{eq3.2}
g_{uu}\, du^2= \mathcal{V}^{-1}\, \mathcal{D}^{2}_{HS}\left[\widetilde{\Psi}(u),\, \widetilde{\Psi}(u+du) \right],
\eeq
with $\mathcal{V}$ as a normalization constant. In this framework, the metric $g_{uu}$ provides a natural means of measuring distances along paths in the space parametized by $u$.

To illustrate this setting, let us take the free boson theory as an example. The Hilbert space of the theory consists of a direct product of sectors, each with fixed momentum $k$, and $\mathcal{V}=\int dk\, \Gamma(ke^{u}/\Lambda)$. 
In addition, for $k\leq \Lambda e^{-u}$, $\Phi_k(u)=\Phi(u) \in \mathbb{R}$ and henceforth, the overlap between two coherent states $|\widetilde{\Psi}(u + du)\rangle$ and $|\widetilde{\Psi}(u)\rangle$ (assuming that $\Phi(u)$ smoothly changes as $u$ varies, i.e, $\Phi(u+du) \approx \Phi(u) + \partial_u \Phi(u) du$), reads as,
\beq
\label{eq3.4}
|\langle \widetilde{\Psi}(u)|\widetilde{\Psi}(u +du)\rangle|^2 = \exp\left[ -\mathcal{V}\, \left(\partial_u\, \Phi(u)\right)^2\, du^2\right] = \exp \left[ -\mathcal{V}\, g_k(u)^2\, du^2\right].
\eeq
Then, making use of eq. (\ref{eq2.7}) one obtains, 
\beq
\label{eq3.4x}
\mathcal{D}^2_{HS}\left[\widetilde{\Psi}(u),\, \widetilde{\Psi}(u+du) \right]  \approx \mathcal{V}\, g_k(u)^2 \, du^2 = \frac{\mathcal{V}}{4}\, \left[\, \partial_u\, S_k(u)\, \right]^2\, du^2.
\eeq
Substituting this result into eq. (\ref{eq3.2}) yields,
\beq
\label{eq3.5}
g_{uu}(u) = \frac{1}{4}\, \left[ (\partial_u\, S_k(u))\, (\partial_u\, S_k(u)) \right],
\eeq 
which explicitly connects the $g_{uu}$ component of the cMERA metric with the entanglement generated at each step of the process. Regarding eqs.(\ref{eq2.7},\, \ref{eq2.8}), $g_{uu}(u)$ explicitly reads as
\beq
\label{eq3.5i}
g_{uu}(u)=\frac{1}{4}\, \frac{e^{-4u}}{\left(e^{-2u} + \bar{m}^2 \right)^2 },
\eeq
with $\bar{m} = m/\Lambda \ll 1$. 

In the holographic dual interpretation of cMERA, for the massless case ($\bar{m}=0$) $g_{uu}(u) = 1/4$ so, eq.(\ref{eq3.0}) would  refer to a pure AdS space. On the other hand, when $\bar{m} \neq 0$, the AdS geometry remains ($g_{uu}(u) \approx 1/4$) for small values of $u$ while it asymptotically vanishes ($g_{uu}(u) \to 0$) for $u \gg -\log \bar{m}\equiv u_{IR}$. In the information theoretic interpretation, this amounts to $\partial_u S_k(u)$ vanishing at a different scale $u_{IR}$ which increases as $m$ decreases. This scale really indicates that the renormalization process has reached the state $|\Psi_{IR}\rangle$.

The emergent cMERA geometrical structures discussed above, happen to be a realization of the recently proposed Surface/State correspondence \cite{prog6, new_taka_cMERA}. The correspondence assigns a dual quantum state $|\Psi(\Sigma)\rangle$ to each space-like surface $\Sigma$ of a dual gravitational theory. This provides a generalized notion of holography as the proposal does not rely on the existence of boundaries in gravitational spacetimes. Essential to this duality is the concept of effective entropy $S_{\rm eff}(\Sigma)$, which amounts to the $\log$ of the effective dimension of the the Hilbert space associated to $\Sigma$. In the discrete version of MERA, $S_{\rm eff}(\Sigma_u)$ measures the number of links of the network which intersect with the surface $\Sigma_u$ given by a fixed scale $u$. 
While the information metric in cMERA was computed in \cite{taka1} for surfaces $\Sigma_u$ to yield $g_{uu}\sim S_{\rm eff}(\Sigma_u)$, here in like manner to \cite{taka1, prog6} (i.e, using the quantum distance between two infinitesimally close $|\Psi(\Sigma)\rangle$ quantum states), we have obtained a geometrical description of cMERA in terms of the entanglement flow along the tensor network, formulated as the left-right entanglement between modes at each length scale $u$ (eq.(\ref{eq3.5})). In the light of these results, one might argue, at least for the cases where cMERA may be casted in terms of coherent states, that the entanglement flow along cMERA turns out to be an effective way to compute $S_{\rm eff}(\Sigma_u)$. 

It is important to note that $ S_{\rm eff}(\Sigma_u)$ cannot be trivially related with the holographic entanglement entropy $S(A)$ of an arbitrary spatial bipartion (which amounts to an arbitrary Hilbert space decomposition different from the left-right moving mode decomposition used in this paper). Even so, in \cite{taka1} it has been shown that $S_{\rm eff}(\Sigma_u)$ directly relates to $S(A)$ when $A$ is half of the space (i.e, as one considers the equal bipartion of the total space). In this case, the holographic entanglement entropy $S(A)$ reads as,
\beq
\label{eq5.3ii}
S(A) \propto \int_{0}^{u_{IR}}\, \sqrt{g_{uu}}\, du \sim \int_{0}^{u_{IR}}\, \sqrt{S_{\rm eff}(\Sigma_u)}\, du.
\eeq

\subsection*{Geometry Fluctuations and Cram\'er-Rao bound}
\label{CramerRao:bound}
An interesting corollary may be obtained from Eq.(\ref{eq3.5}). Let us consider an observer (with density matrix $\widetilde{\rho} \equiv |\widetilde{\Psi}(u)\rangle \langle \widetilde{\Psi}(u)|$ ), wishing to estimate the value of the radial coordinate $u$ through a measurement of the position operator $\mathit{\widehat{X}_u}$ such that $\langle \mathit{\widehat{X}_u} \rangle = {\rm Tr}\left(\, \widetilde{\rho}\, \mathit{\widehat{X}_u}\, \right) $. An important result in information theory known as the Cram\'er-Rao bound \cite{amari}, establishes that the lowest bound for $\langle (\delta u)^2\rangle = {\rm Tr}\left(\, \widetilde{\rho}\, (\mathit{\widehat{X}_u}-u)^2\, \right)$ is given by,
\beq
\label{eq3.5x}
\langle (\delta u)^2\rangle \geq \frac{1}{4\, g_{uu}}.
\eeq
This states that the larger are the changes in the probability distributions along the $u$-coordinate (measured by the Fisher metric), the better are the estimations for the value of this coordinate. For the bosonic theory, the bound reads as,
\beq
\label{eq3.5y}
\langle (\delta u)^2\rangle \geq \frac{1}{\left[\, \partial_u\, S_k(u)\, \right]^{2}}.
\eeq 
In the massless limit of this theory, both $S_k(u)$ and $\partial_u\, S_k(u)$ must be proportional to the central charge $\mathit{C}$ of the theory so, $\langle (\delta u)^2\rangle \sim \mathit{C}^{-2}$. Thus, if one conjectures that for the cMERA construction of theories with large $C$, still holds that the Fisher information metric $g_{uu} \propto \left[\, \partial_u\, S_k(u)\, \right]^{2}$ (and hence eq.(\ref{eq3.5y})), then, as a result, one gets that the estimation error $\langle (\delta u)^2\rangle $ would become largely suppressed for those theories. This seems to conform to the emergence of classical geometries in the large $\mathit{C}$ limit, as stated by the AdS/CFT correspondence \citep{prog6}.

\subsection*{cMERA gauge invariance and Fisher metric}
\label{gauge_CMERA}
Finally, we show how certain \textit{gauge invariance} of the cMERA flow directly reflects on the invariance of $g_{uu}$. To this end, let us first note that the cMERA \textit{evolution} operator
\beq
\label{eq3.5z}
U(u_{*}|u_{IR})=\exp\left(-i\int_{u_{IR}}^{u_{*}}\, \widetilde{K}(\hat{u})\, d\hat{u}\right),
\eeq
can be written as,
\barray
\label{eq3.6}
U(u_{*}|u_{IR}) = U(u_{*}|u_{*}+\delta_u)\cdots U(u-\delta_u|u)\, U(u|u+\delta_u)\cdots U(u_{IR}-\delta_u|u_{IR}).
\earray
Here, $\delta_u=du$ and
\beq
\label{eq3.6i}
U(u|u+\delta_u)=\exp\left( -i\widetilde{K}(u)\, \delta_u\right),
\eeq
corresponds to an infinitesimal layer of cMERA. It happens that the operator $U(u_{*}|u_{IR})$ remains invariant if one inserts the product $G^{\dagger}(u)\, G(u)$ of a unitary scale-dependent gauge transformation $G(u)$ and its inverse $G^{\dagger}(u)$ in between any two layers
\beq
\label{eq3.7}
U(u-\delta_u|u)\, G^{\dagger}(u)\, G(u)\, U(u|u+\delta_u).
\eeq
In addition, one must impose that
\beq
\label{eq3.7i}
G(u_{IR})|\Psi_{IR}\rangle =|\Psi_{IR}\rangle,
\eeq 
to guarantee that  $|\widetilde{\Psi}(u)\rangle$ remains invariant. The gauge transformed layer operator 
\beq
\label{eq3.7ii}
\bar{U}(u|u+du)= G(u)\, U(u|u+du)\, G^{\dagger}(u + du),
\eeq
up to first order in $du$ reads as,
\beq
\label{eq3.8}
\bar{U}(u|u+du)= \exp\left[ -i du\, \left(  G(u)\, \widetilde{K}(u) G^{\dagger}(u) + i\, G(u)\,\partial_u\, G^{\dagger}(u)\right)\right].
\eeq
Thus, under a gauge transformation $G(u)$, the entanglement generator of the cMERA flow $\widetilde{K}(u)$ transforms as,
\beq
\label{eq3.9}
\widetilde{K}^{'}(u) = G(u)\, \widetilde{K}(u)\, G^{\dagger}(u) + i\, G(u)\,\partial_u\, G^{\dagger}(u).
\eeq

Here we are interested in the class of gauge transformations which leaves the Fisher information metric $g_{uu}(u)$ in (\ref{eq3.5}) invariant. Formally, the Fisher metric is defined through,
\beq
\label{eq3.10}
g_{uu}=\langle \partial_u \widetilde{\Psi}(u) |\partial_u \widetilde{\Psi}(u)\rangle - \langle \partial_u \widetilde{\Psi}(u) | \widetilde{\Psi}(u)\rangle \langle \widetilde{\Psi}(u) |\partial_u \widetilde{\Psi}(u)\rangle,
\eeq
where we have used the compressed notation
\beq
\label{eq3.10i}
|\partial_u \widetilde{\Psi}(u)\rangle\equiv \partial_u\, |\widetilde{\Psi}(u)\rangle = -i\widetilde{K}(u)|\widetilde{\Psi}(u)\rangle.
\eeq
With this, $g_ {uu}$ can be written as the variance of $\widetilde{K}(u)$,
\beq
\label{eq3.11}
g_{uu}=\langle \widetilde{\Psi}(u) |\widetilde{K}(u)^2| \widetilde{\Psi}(u)\rangle - \langle  \widetilde{\Psi}(u) | \widetilde{K}(u)|\widetilde{\Psi}(u)\rangle^2 .
\eeq
Now, regarding eq.(\ref{eq3.9}), it is clear that $g_{uu}$ will remain invariant under a gauge transformation $G(u)$ of the cMERA flow provided that $\widetilde{K}(u) =   \widetilde{K}^{'}(u)$. In this sense, let us focus on the gauge transformations $G(u)=\exp \left( i\epsilon\, \mathcal{O}(u) \right)$ generated by a self-adjoint scale-dependent local operator $\mathcal{O}(u)$, with $\epsilon$ being a small parameter. Under these assumptions, the transformation in (\ref{eq3.9}) can be written as,
\beq
\label{eq3.12}
\widetilde{K}^{'}(u) =\widetilde{K}(u) + \epsilon\, \nabla_u\,  \mathcal{O}(u),   
\eeq
with
\beq
\label{eq3.12x}
\nabla_u\, \mathcal{O}(u) =  \partial_u\, \mathcal{O}(u)-i\left[\widetilde{K}(u),\, \mathcal{O}(u) \right].
\eeq
Here, we note that the equation which defines the cMERA flow for an operator such as $\mathcal{O}(u)$, is similar to the equation of motion of an operator in the Heisenberg picture with respect to the $u$-dependent \textit{Hamiltonian} $\widetilde{K}(u)$ \cite{cMERA}. This equation reads as,
\beq
\label{eq3.13}
\partial_u\, \mathcal{O}(u)=i\left[\widetilde{K}(u),\, \mathcal{O}(u) \right]. 
\eeq
which implies that,
\beq
\label{eq3.14}
\nabla_u\,  \mathcal{O}(u)=0,
\eeq
and thus, $\widetilde{K}^{'}(u)=\widetilde{K}(u)$. Regarding eq.(\ref{eq3.11}), this condition assures the invariance of $g_{uu}$ for this class of gauge transformations on the cMERA flow.

\section{Conclusions}
\label{conclusions}
We have explicitly shown how the information metric emerging from a static cMERA state amounts to the differential generation of entanglement entropy along the renormalization group flow. We also characterized a class of gauge transformations of this flow which leaves the metric invariant. The results have been derived  only for free gaussian theories so, it would be desirable to check if these results possess some useful generalizations in the case of interacting theories. In this sense, the ground state and correlation functions in a theory of interacting (1+1) bosons have been computed by means of a continuous version of the matrix product state tensor network (cMPS) \cite{cMPS}. Nevertheless, in these tensor networks, the entanglement structures needed to build a state deviates from the multiscale analysis carried out by cMERA. Further investigations might also address non-stationary settings such as quantum quenches. Entanglement renormalization deals with time dependent states \cite{taka1} but tackling space and time on quite different grounds. It is worth to investigate if the analysis of the entanglement flow in this states could provide some light in order to formulate a time-dependent cMERA in a covariant way. Finally, it would be also worth to clarify if it is possible to generate cMERA states/cMERA entanglement flows, compatible with information metrics which extremize a (gravitational-like) action functional.

\section*{Acknowledgements}
JMV gratefully thanks Esperanza L\'opez, Emilia da Silva and Germ\'an Sierra for very fruitful discussions and their hospitality at Instituto de F\'isica Te\'orica CSIC-UAM in Madrid. The author also thanks  Tadashi Takayanagi and Bartlomiej Czech for stimulating discussions on holography and tensor networks during the Workshop "Entangle This: Space, Time and Matter". This work has been supported by  Ministerio de Econom\'ia y Competitividad of Spain Project No. FIS2012-30625

\end{document}